\documentclass{aa}                       
\usepackage{psfig}
\begin{document}
\thesaurus{11 (11.09.1, NGC 2146, 11.09.4, 11.11.1, 11.19.3)}
\title{The minor axis outflow of NGC 2146}
\author{A.~Greve\inst{1}
 \and N.~Neininger\inst{2} \and A.~Tarchi\inst{2}
 \and A.~Sievers\inst{3} }
\offprints{A.~Greve}
\institute{Institut de Radio Astronomie Millim\'etrique,
           300 rue de la Piscine, 38406 St.\ Martin d`H\`eres, France
           \and Astronomisches Institut der Universit\"at Bonn, 71 Auf
           dem H\"ugel, 53121 Bonn, Germany \and Instituto
           Radioastronom\'{\i}a Milim\'etrica, Nucleo Central, 7
           Avenida Divina Pastora, 18012 Granada, Spain }
\date{received date; accepted date\\
}
\maketitle
\begin{abstract}
X--ray and optical observations have shown that the supernova explosions and 
stellar winds of the starburst in NGC\,2146 produce an outflow of hot material
along the minor axis. This outflow is expected to have a more or less conical 
shape, on either side of the galactic plane, and cone walls of cooler material
where the outflow is in shock contact with halo gas. We attempt to determine 
the geometry (diameter at the base and opening angle) and the physical 
parameters (velocity and density) of the material in the cone walls from the 
optical emission line and radio observations presented here, and from published
X--ray, radio, and optical observations. We compare the outflow of NGC\,2146 
with the outflow of M\,82.

\end{abstract}

\section{Introduction} 
In analogy to other starburst galaxies (Heckman et al.\ 1990, Leh\-nert $\&$ 
Heckman 1996\,a), the supernova explosions and stellar winds of the strong
star formation activity in NGC\,2146 produce an outflow of hot gas along the 
minor axis. The outflow is collimated in the disk by a ring of dense molecular
gas confining the star formation region, and hence appears at X--ray and 
optical wavelengths under ideal conditions as cones on either side of the 
galactic plane. Such an outflow is especially prominent in galaxies viewed 
edge--on, as in the case of M\,82. The hot gas at a temperature of 10$^{7}$ --
10$^{8}$\,K, seen at X--rays, is confined to the inner part of the cones; the 
gas in the cone walls in shock contact with the halo material is at a 
temperature of $\sim$\,10$^{4}$\,K and emits the [forbidden] recombination 
lines of H$_{\alpha}$, [NII], [SII], etc. The outflowing hot gas may drag cool
gas and dust out of the disk at the inner edge of the molecular ring and 
transports this material, and angular momentum, to a considerable height in 
the halo. The structure of such an outflow is schema\-tically shown by Heckman
et al.\ (1990, their Fig.\,1) and clearly seen in M\,82 at X--ray (Schaaf et 
al.\ 1989, Bregman et al.\ 1995) and optical wavelengths (Axon $\&$ Taylor 
1978, Bland $\&$ Tully 1988, Shopbell $\&$ Bland--Hawthorn 1998, McKeith et al.
1995, Devine $\&$ Bally 1999). Predictions, calculations, and interpretations 
of these outflows under the aspect of energy input, outflow velocity, density 
and temperature of the gas, and height of the outflow were published by 
Chevalier $\&$ Clegg (1985), Tomisaka $\&$ Ikeuchi (1988), Umemura et al. 
(1988), Yokoo et al. (1993), and others.

The X--ray and optical emission of the outflow in NGC 2146 was observed by 
Armus et al. (1995) and Della Ceca et al. (1999). Ultra--compact/ultra--dense 
H\,{\sc ii} regions containing many O--type stars, and radio supernovae and 
supernova remnants were recently detected in the starburst region in a
combination of MERLIN and VLA observations by Tarchi et al. (2000).

\begin{figure}
\psfig{figure=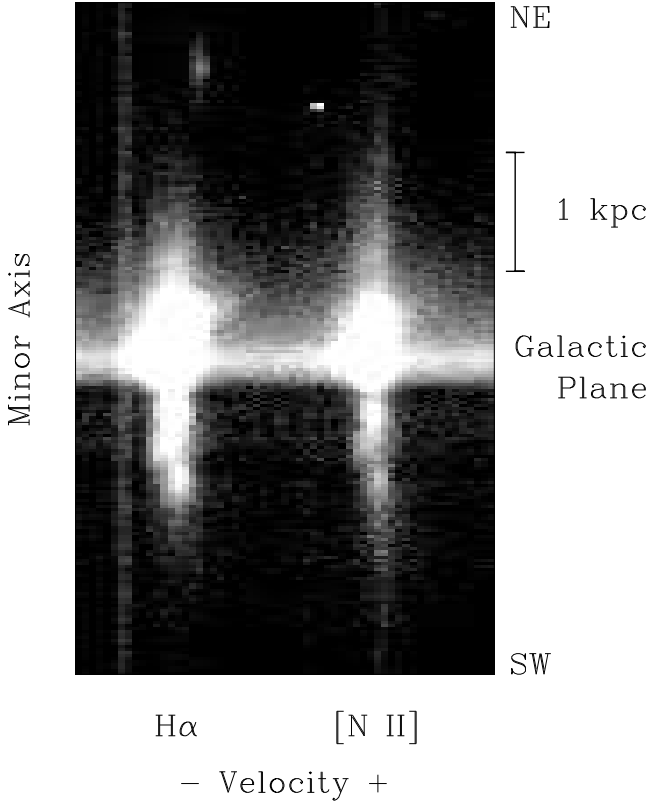,height=11.0cm,angle=0} 
\caption[]{H$_{\alpha}$\,6563\,\AA\ and [NII]\,6583\,\AA\ measured along the 
minor axis (pa = 50$\degr$) of NGC\,2146. The [vertical] straight line at the 
very left is a sky emission line. The [horizontal] continuum emission comes 
from a region near the center of the galactic plane. Note the increased 
emission from an H\,{\sc ii} region complex at $\sim$\,2.5\,kpc NE above the 
galactic plane (see Fig.\,4 in Young et al., 1988).}
\end{figure}

Optical spectroscopy observations with the slit oriented along the minor axis 
show under favourable geo\-metrical conditions characteristic velocity--split
emission lines (M\,82); conversely, measurement of the line splitting allows 
a determination of the outflow geo\-metry and the outflow veloci\-ty. 
Unfortunately, for reasons to be explained below, in NGC\,2146 there is no 
very clear detection of line splitting and hence the geo\-metry and the 
velocity of the outflow is not well known, although the galaxy is seen tilted 
from edge--on by only 25 -- 30\,$\degr$. From a combination of the observations
of this paper (optical spectroscopy, CO line interferometer data) and published
X--ray, radio, and optical measurements we attempt to determine the geo\-metry
and the physical parameters of the outflow cones: i.e.\ the dia\-meter ({\it 
b}) of the base in the galactic plane, the opening angle ($\Theta$), the 
outflow velocity ({\it V}), and the (electron) density (n$_{\rm e}$) of the 
material in the cone walls. We compare the outflow of NGC\,2146 with the well 
studied outflow of M\,82.

\section{The starburst galaxy NGC\,2146}
NGC\,2146, at a distance of 14.5\,Mpc (Benvenuti et al.\ 1975, for H$_{0}$ = 75;
1$''$ is equivalent to 70\,pc), is experiencing a starburst in the central 
region of $\sim$\,2.5\,kpc dia\-meter. The molecular gas (CO) in this region is 
concentrated in a warped disk and a ring of similar dimension (Fig.\,10 below).
Since NGC\,2146 has no companion galaxy (at least when using the criteria of 
spatial and velocity coincidence; Fisher $\&$ Tully 1976), it has been proposed
that the starburst is triggered by a far evolved merger (Jackson $\&$ Ho 1988,
Young et al.\ 1988, Hutchings et al.\ 1990, Taramopoulos et al. 2000). However,
the meager kinematic and material signature of the merger is far from 
convincing.

The galaxy is seen nearly edge--on\footnote{face--on: {\it i} = 0$\degr$.} 
at the inclination \mbox{{\it i} $\approx$\,60 -- 65\,$\degr$} so that the
minor axis is tilted by $\epsilon$ = 90$\degr$ -- {\it i} $\approx$ 25 --
30$\degr$ out of the plane of the sky; the position angle of the major axis is
pa $\approx$ 145$\degr$; the morphology is disturbed and distorted; the center
is moderately obscured by A$_{\rm V}$ $\approx$ 4\,--\,7\,mag extinction 
(Benvenuti et al.\ 1975, Burbidge et al.\ 1959, Smith et al.\ 1995). Compared 
to M\,82, the five times larger distance of NGC\,2146 makes (detailed) 
observations more difficult. 

\section{Optical observations along the minor axis}
In long--slit spectroscopy observations, the position--veloci\-ty signature of
the outflow is determined by the geo\-metry of the outflow cones ({\it b}, 
$\Theta$, and the skewness of the cones with respect to the galactic plane; 
see Fig.\,9), the velocity of the outflow, the orientation of the galaxy, and 
the orientation of the slit. Observations with the slit oriented along the 
minor axis are expected to show velocity--split emission lines, similar to 
those seen in the outflow of M\,82 (Mc\-Keith et al.\ 1995, Shopbell $\&$ 
Bland--Hawthorn 1996). Burbidge et al.\ (1959) reported line splitting of 
$\Delta${\it v} $\approx$ 300\,km\,s$^{-1}$ near the center of NGC\,2146. 
Hutchings et al.\ (1990) mention line splitting, though without further detail.
Armus et al.\ (1995) report that the emission of the outflow is blue--shifted 
at the NE side and red--shifted at the SW side with a velocity separation 
of $\sim$\,130\,km\,s$^{-1}$, that line splitting of $\Delta${\it v} $\approx$
150 -- 200\,km\,s$^{-1}$ occurs at 1.5 -- 3\,kpc distance on either side of the
galactic plane, and that the line width increases from 
$\sim$\,100\,km\,s$^{-1}$ to 200 -- 300\,km\,s$^{-1}$ at a height of
$\sim$\,1\,kpc. Line splitting was not seen by Benvenuti et al.\ (1975) and 
Young et al.\ (1988).

\begin{figure}
\psfig{figure=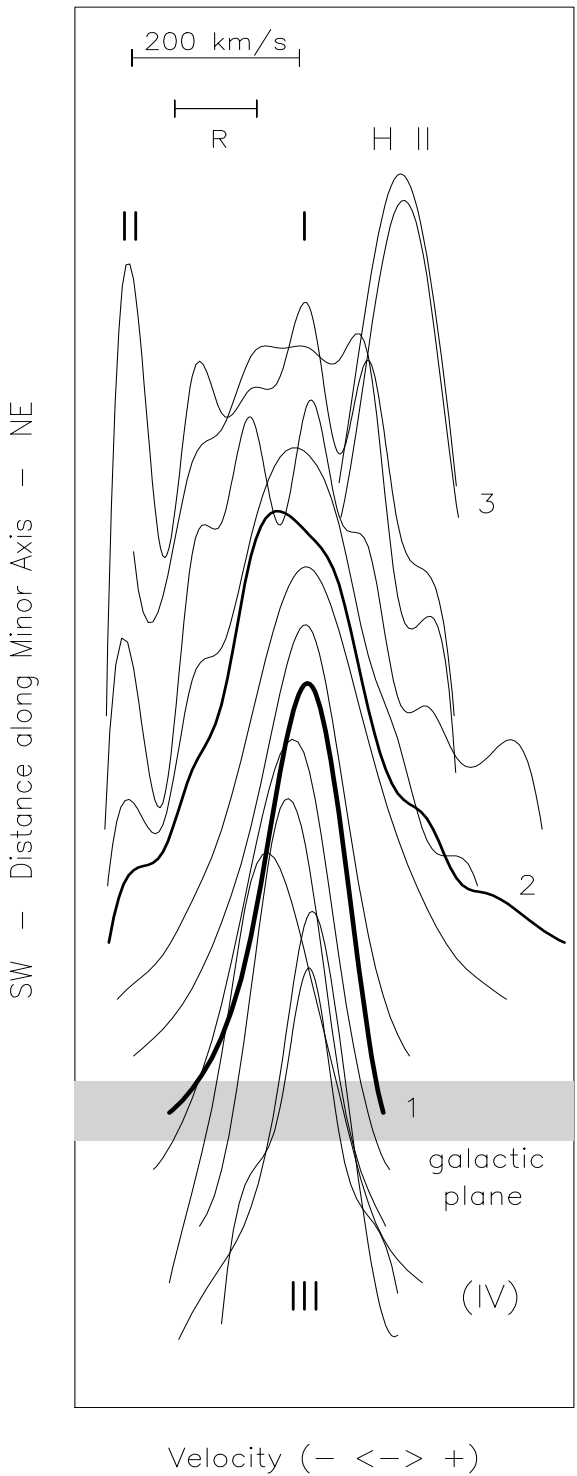,width=7.5cm,height=19.5cm,angle=0} 
\caption[]{H$_{\alpha}$ velocity tracings at various heights along the minor 
axis below (SW) and above (NE) the galactic plane illustrating the differences
in profile shape. The components I -- (IV) of the outflow cone walls are 
indicated at their appropriate location along the velocity axis (see also 
Fig.\,9). \mbox{Profile 1} is extracted at the galactic plane; 
\mbox{profile 2} at $\sim$\,0.5\,kpc NE above the galactic plane has the 
largest width; \mbox{profile 3} is the H\,{\sc ii} region at the NE (see 
Fig.\,1). The velocity scale and the resolution of the spectrogram (R) is 
inserted. Intensities are not to scale.}
\end{figure}

With the 4.2\,m William Herschel telescope (La Palma) we have obtained a 
spectrogram, containing the H$_{\alpha}$\,6563\,\AA, [NII]\,6548,\,6583\,\AA, 
and [SII]\,6716,\,6731\,\AA\ emission lines, with the slit oriented along the 
minor axis (pa = 50$\degr$) and passing through the eye--estimated 'center' of
the galaxy (i.e. the brightest central spot). The velocity resolution 
determined from sky lines is $\sim$\,100\,km\,s$^{-1}$ (FWHM of a Gaussian 
profile); velocity structures of the line profiles can be determined with a
precision of $\sim$\,30\,km\,s$^{-1}$. The H$_{\alpha}$ and
[NII]\,6583\,\AA\ line are reproduced in Fig.\,1. The picture shows strong line
(gas) and continuum (stars) emission in the galactic disk of $\sim$\,0.5\,kpc 
thickness, break--out of the gas from the disk, and faint emission extending 
into the halo to a height of $\sim$\,3\,kpc (limited by the length of the slit
and the exposure time of 1\,800\,s). The picture shows noticeable H$_{\alpha}$
emission at a height of $\sim$\,2.5\,kpc to the NE. This emission comes from 
an object in the string of H\,{\sc ii} regions having a velocity of 
\mbox{$\sim$ +\,100\,km\,s$^{-1}$} with respect to the center of the galaxy. 
This, or a similar H\,{\sc ii} region, is also seen in the spectrogram taken 
along the minor axis by Armus et al. (1995, their Fig.\,8). The string of 
H\,{\sc ii} regions is clearly seen in the pictures published by Young et 
al.\ (1988).

\section{Outflow velocities}
From the H$_{\alpha}$ line (Fig.\,1) we have extracted velocity tracings at 
various heights below and above the galactic plane. These tracings, normalized
to the individual peak emission (see also Fig.\,4\,a below), are shown in 
Fig.\,2 and are interpreted together with Fig.\,9 below which explains the 
proposed geometry of the outflow and the velocity components I -- IV. 
Contrary\footnote{part of the observed spectral differences are certainly due 
to different slit positions.} to the observation by Armus et al.\ (1995) we do
not find a systematic blue/red--shift of the lines at the NE/SW side of the 
galactic plane (see the straight [NII] line in Fig.\,1). Also contrary to their
observation, the profiles at the SW side of the galactic plane do not show 
line splitting, but only component III. The profiles at the NE side show 
primarily component I, but also some line splitting as illustrated in Fig.\,3.
The intensity of the tentatively identified component II is {\it very weak}, 
however, the appearance of component II in several tracings covering a distance
of $\sim$\,1\,kpc NE along the minor axis (between $\sim$\,0.5\,kpc and 
$\sim$\,1.5\,kpc) gives confidence of its existence. At $\sim$\,1.5\,kpc NE 
above the galactic plane the splitting between component I (at positive 
velocities) and component II (at negative velocities) is $\Delta${\it v} 
$\approx$ 200\,km\,s$^{-1}$; a similar value was found by Armus et al. (1995).
The width of 300 -- 400\,km\,s$^{-1}$ of profile 2 (Fig.\,2) at 
$\sim$\,0.5\,kpc NE above the galactic plane is probably due to acceleration 
of the outflow when breaking out of the galactic plane, similar as seen in
M\,82 (McKeith et al. 1995, their Fig.\,2). The component IV at the SW is not 
seen. We interpret the observation that component I and III appear at {\it 
nearly} the same and constant velocity (Fig.\,2) by the fact that the 
line--of--sight is {\it nearly} perpendicular to the corresponding cone walls.
For a similar situation in M\,82, where all components are observed, see 
Mc\-Keith et al. (1995).

\begin{figure}
\psfig{figure=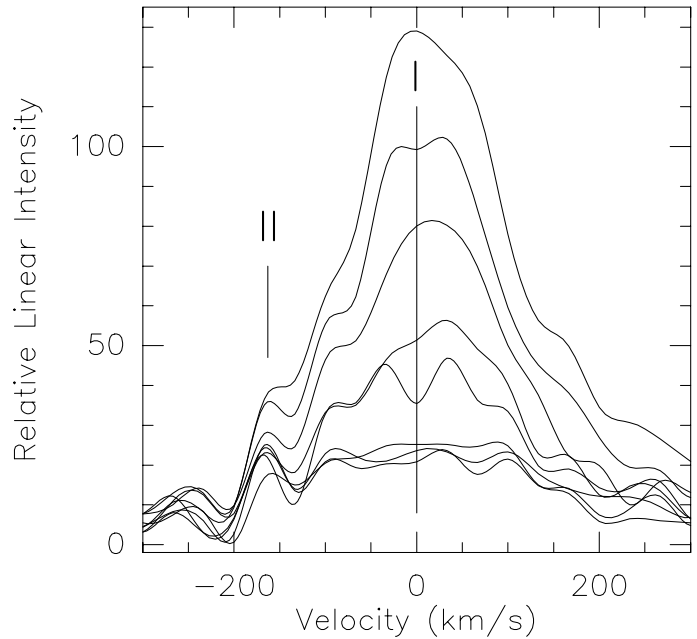,width=9.0cm,angle=0} 
\caption[]{H$_{\alpha}$ velocity tracings between $\sim$\,0.5\,kpc and
$\sim$\,1.5\,kpc above the galactic plane (NE) where component I and component
II are seen. [The noise in the tracings is visible at velocities below
--\,200\,km\,s$^{-1}$.]} 
\end{figure}

\begin{figure}
\psfig{figure=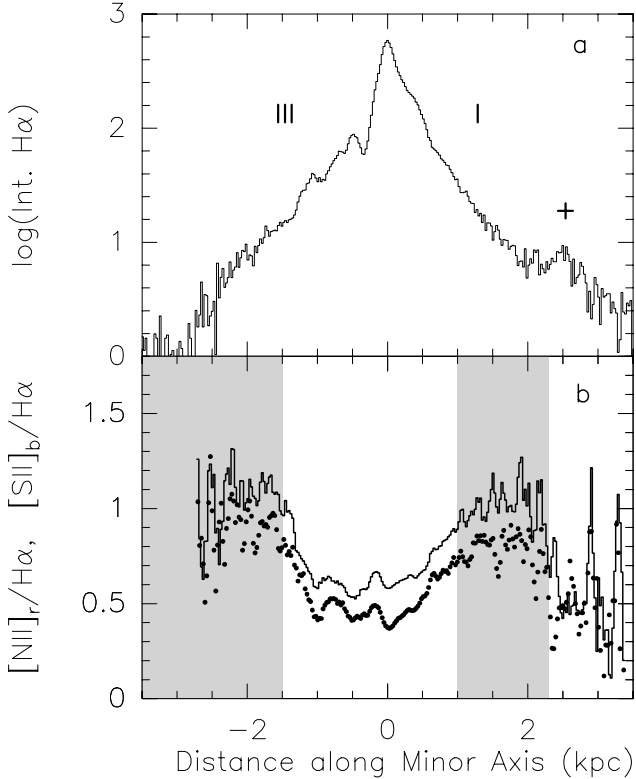,width=8.3cm,angle=0}
\caption[]{{\bf (a)} H$_{\alpha}$ emission (logarithm of relative units) 
measured along the minor axis. Gas of the minor axis outflow is indicated by
the shaded areas.  Outside the galactic disk, the intensities at the SW and NE
side are primarily emission of component III and component I (see 
Fig.\,2,\,3\,$\&$\,9). The mark (+) indicates the H\,{\sc ii} region at 
$\sim$\,2.5\,kpc NE above the disk (see Fig.\,1). \mbox{{\bf (b)} Ratios}
[NII]$_{\rm r}$/H$_{\alpha}$ (line) and [SII]$_{\rm b}$/H$_{\alpha}$ (dots) 
which indicate thermally excited gas \mbox{($\approx$ 0.5)} and gas with a 
contribution of shock excited gas ($\approx$\,1). SW is to the left, NE to the
right.}
\end{figure}

\begin{figure}
\psfig{figure=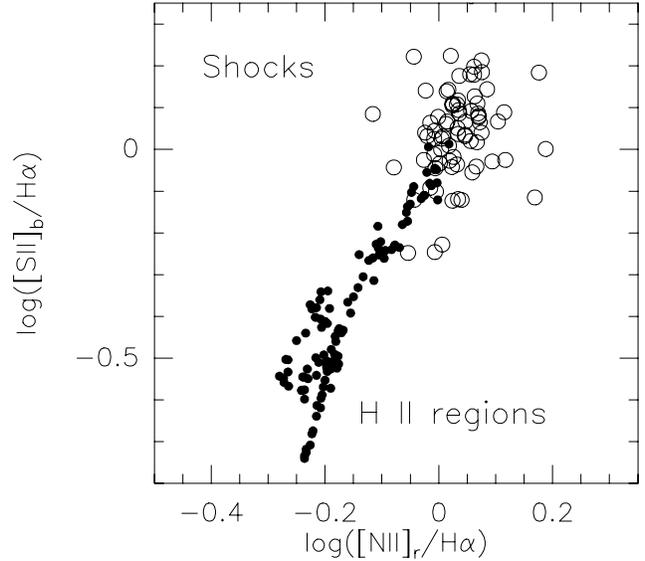,width=8.3cm,angle=0}
\caption[]{Line ratio diagram of gas measured along the minor axis. Open 
circles: outflow gas at a height of --3\,kpc $\la$ {\it z} 
\mbox{$\la$ --\,1.5\,kpc} and 1\,kpc $\la$ {\it z} $\la$ 2.3\,kpc below 
and above the galactic plane (shaded region in Fig.\,4\,b) containing a 
contribution of shock excited gas. Dots: gas at a height of --1.5\,kpc $\la$ 
{\it z} $\la$ 1.2\,kpc containing thermally excited gas, primarily in the 
galactic disk. For similar diagrams of Diffuse Interstellar Gas in galaxies 
see Rand (1998) and Leh\-nert $\& $ Heckman (1996\,b).}
\end{figure}

\section{Line ratios and density}
The [NII]\,6583/H$_{\alpha}$ $\equiv$ [NII]$_{\rm r}$/H$_{\alpha}$ and
[SII]\,6716 /H$_{\alpha}$ $\equiv$ [SII]$_{\rm b}$/ H$_{\alpha}$ line ratios 
are used to investigate shock excitation of the ionized gas; the 
[SII]\,6716/6731 line ratio is used to derive the electron density n$_{\rm e}$
(for T$_{\rm e}$ = 10$^{4}$\,K) (Osterbrock 1989, Lehnert $\&$ Heckman 1996\,b).
These line ratios need no correction for differential extinction, but the 
location in the galaxy to which they refer depends on the line--of--sight 
absorption (probably not exceeding A$_{\rm V}$ $\approx$ 4\,--\,7\,mag at 
the center). The H$_{\alpha}$ emission, and the H$_{\alpha}$ line profiles 
extracted in the galactic disk, may need a correction for underlying stellar 
H$_{\alpha}$ absorption. We have no data of the stellar line, and neglect the 
influence in the following comparative study.

We have extracted from the spectrogram of Fig.\,1 the line ratios
[NII]$_{\rm r}$/H$_{\alpha}$ and [SII]$_{\rm b}$/H$_{\alpha}$ shown in 
Fig.\,4\,b. The values of this figure agree with the observations by Armus et 
al.\ (1995, their Fig.\,6). At the center of the galaxy, and to a height of 
0.5 -- 1\,kpc above the disk, we find values of $\sim$\,0.5 as typical for 
thermally excited ionized gas in H\,{\sc ii} regions. This is confirmed by the
clear signature of the H\,{\sc ii} region gas at $\sim$\,2.5\,kpc NE above the
disk. Between $\sim$\,1\,kpc and $\sim$\,2\,kpc above the disk the line ratios
gradually increase to $\sim$\,1 and remain at this value to a height of 3\,kpc,
indicating ionized gas with a noticeable contribution of shock excited gas 
(Rand 1998, Martin 1998, Wang et al.\ 1999, Lehnert $\&$ Heckman 1996\,b, 
Reynolds et al.\ 1999), as expected to be present in the walls of the outflow 
cones\footnote{The line ratios can, in principle, only be calculated for the 
individual components I -- IV (for M\,82 see Mc\-Keith et al.\ 1995). We 
explain below (Fig.\,9) the geometry of the outflow cones and the reason of 
using the line ratios of the velocity--integrated emission NE and SW of the 
disk.}. The additional diagnostic dia\-gram Fig.\,5 confirms that the gas at a
height of --\,3\,kpc $\la$ {\it z} $\la$ --\,1.5\,kpc and 1\,kpc $\la$ {\it z}
$\la$ 2.3\,kpc above the plane, which in essence is gas of the outflow, 
contains a contribution of shock excited gas, while the gas at lower heights 
is thermally excited.

The line ratio [SII]\,6716/6731 (see footnote 3) is 0.90 -- 0.95 in the galactic 
disk, $\sim$\,1.1 at a height of 1.5 -- 2\,kpc above the plane, and $\sim$\,1.4
(the low density value) at 2.5 -- 3\,kpc height. These observed values and the
corresponding electron densities are given in \mbox{Table 1.} The values seem 
to indicate a decrease of the electron density in the cone walls with height 
{\it z} which, when approximated by a {\it z}$^{-2}$ dependence of the electron
density as found for M\,82 (McKeith et al. 1995), gives
\begin{equation}
{\rm n}_{\rm e}\ {\approx}\ 600\,[{\rm cm}^{-3}]\,{\rm (z/kpc)}^{-2}
\end{equation}
The values at $\sim$\,1.5\,kpc $\la$ {\it z} are uncertain and need additional
observations. 

\begin{table}
\caption[]{Electron density (n$_{\rm e}$, for T = 10$^{4}$\,K) of the ionized 
gas in the Galactic Disk and the Cone Walls of the outflow in NGC\,2146.}
\begin{center}\small
\begin{tabular}{cccc}
\hline
 & Height {\it z}  & [SII]6716/6731 & n$_{\rm e}$ 
 \\
 & [kpc]     &  (observed)    & [cm$^{-3}$]  \\
\hline
Galactic Disk & 0 -- 0.5 & 0.90 -- 0.95  & $\sim$\,600  \\
Cone Walls & 1.5 & 1.1 & 300 \\
 $''$ & 2.5 & 1.4 & $\la$\,100 \\
\hline
\end{tabular}
\end{center}
\end{table} 

\section{Radio observations}
\subsection{Plateau de Bure CO interferometer observations}
We have used the PdB interferometer in a 7--mosaic observation to map the 
starburst region in the optically thick \mbox{$^{12}$CO\,(1--0)} line and the 
optically thin $^{13}$CO\,(1--0) line (for the optical thickness see Xie et 
al., 1995). For a description of the PdB observations and the results see 
Greve et al. (in prep.). In these observations, and the $^{12}$CO\,(1--0) 
observation by Young et al. (1988), we believe that there is evidence of some
molecular material which is dragged--out along the minor axis. In other 
starburst galaxies, for instance M\,82, a similar drag--out of molecular 
material (CO) into the halo was already proposed in 1987 by Nakai et al. 
(1987); a possible outflow of dust was recently suggested by Thuma et al.
(2000). A molecular outflow along the minor axis of NGC\,253, though only to 
a height of $\sim$\,150\,pc above the galactic plane, was recently observed by 
Garc\'{\i}a--Burillo et al.\ (2000).

\begin{figure*}
\psfig{figure=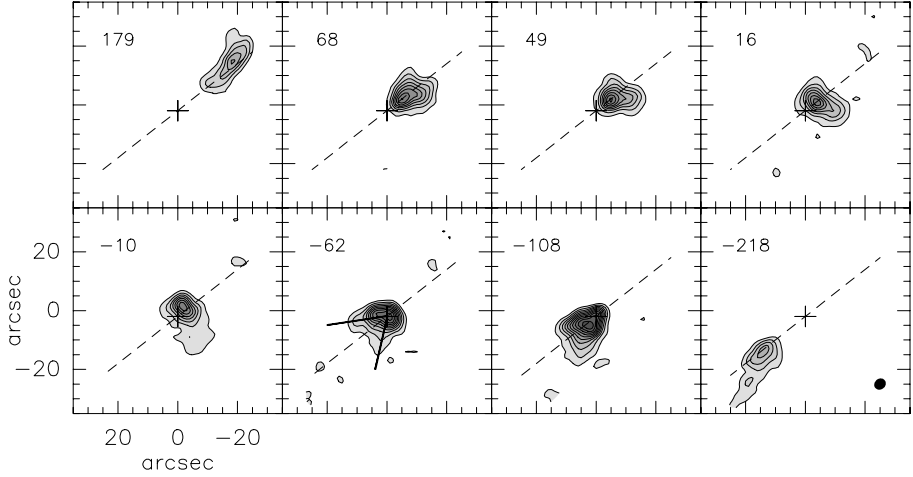,width=12.0cm,angle=-90}
\caption[]{Selected velocity channel maps (with the systemic velocity of 
850\,km\,s$^{-1}$ subtracted) showing {\bf 12}\,CO\,(1--0) material assumed to
be dragged--out of the galactic plane. The method of deriving the position and
direction of the outflow is illustrated for the channel map at
\mbox{--\,62\,km\,s$^{-1}$.} The dashed line is the galactic plane; the solid 
lines indicate the direction of the outflow. The solid dot in the lower
box is the synthesized beam of 4.1$''$$\times$\,3.6$''$ 
[290\,pc$\times$250\,pc]. Contour levels from 0.05 to 0.8 Jy/beam, by 0.075. 
The coordinates are centered (cross) at \mbox{6$^{\rm h}$ 18$^{\rm h}$ 
38.6$^{\rm s}$,} 78$\degr$ 21$'$ 24.0$''$ (2000). N is up, E is left.}
\end{figure*}

\begin{figure*}
\psfig{figure=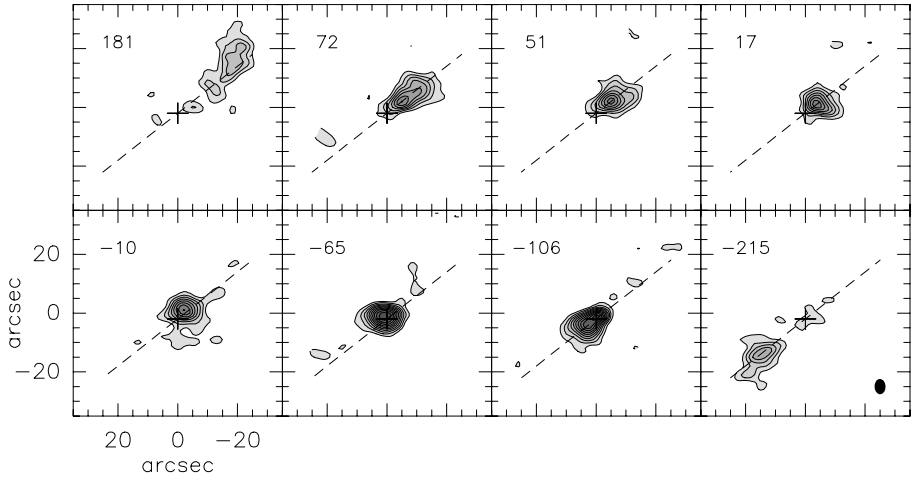,width=12.0cm,angle=-90}
\caption[]{Same as Fig.\,6, but for the approximately {\it ten times weaker} 
{\bf 13}\,CO \mbox{(1--0)} line. Contour levels from 0.005 to 0.1 Jy/beam, by 
0.005. The synthesized beam is 5.2$''$$\times$3.6$''$ [360\,pc$\times$250\,pc].}
\end{figure*}

\subsection{Dragged--out molecular material}
The velocity channel maps of the interferometer measurements by Young et al. 
(1988) and Greve et al.\ (in prep.) cover the starburst region in $^{12}$CO 
and $^{13}$CO and show the bulk of molecular material of the ring and warp and
'fingers' of emission which extend more or less skew out of the galactic plane,
similar to 'chimneys' or 'galactic fountains' seen at radio and optical 
wavelengths. We believe that these outflow features provide evidence of gas 
and dust being dragged--out of the disk at the inner peri\-phery of the 
molecular ring by the superwind of the starburst. This material gives 
information on the geometry of the outflow.

For several selected velocity channels of the \mbox{$^{12}$CO\,(1--0)} and 
$^{13}$CO\,(1--0) line, observed at PdB, we show in Fig.\,6 and Fig.\,7 the 
features which we interpret to indicate outflow. The method of deriving from 
these measurements the position and direction of the outflow (with respect to 
the galactic plane) is illustrated for the $^{12}$CO\,(1--0) channel map at 
--\,62\,km\,s$^{-1}$ (Fig.\,6). In this figure the dashed line is the galactic
plane. The skew solid lines, above and below the galactic plane, indicate 
at 0.3 -- 0.5\,kpc height a component of dragged--out molecular material
and the direction of the outflow. The base of this outflow is near the
intersection of the solid lines with the galactic plane; the relative height 
of the outflow is the distance of a selected lowest contour measured in 
the direction of the outflow. We have investigated corresponding channel
maps of the $^{12}$CO\,(1--0) (Fig.\,6) and the \mbox{$^{13}$CO\,(1--0)}
(Fig.\,7) line in order to assess that the observed features are not 
produced by an optical depth effect. The correspondence is good, taking 
into account the $\sim$\,10 times weaker intensity of the $^{13}$CO line.
This and similar evidence from other observations is used in Fig.\,8 for a 
determination of the outflow geometry.

\section{Derivation of the outflow geometry}
In analogy to similar phenomena observed in M\,82, we believe that the {\it 
ensemble} of several emission features provides also in NGC\,2146 evidence of 
the outflow and information of the outflow geometry:

\begin{figure}
\psfig{figure=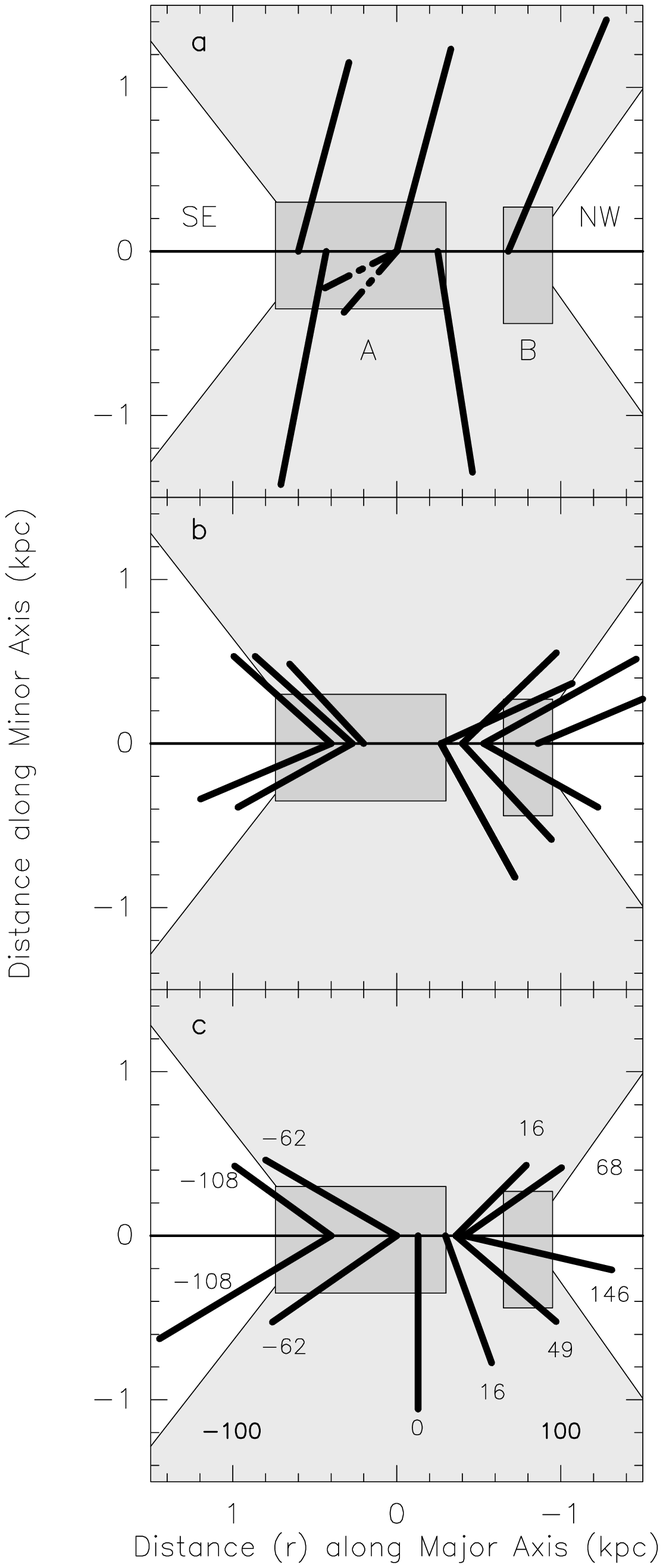,width=7.4cm,angle=0}
\caption[]{Derivation of the outflow base ({\it b}) and cone opening angle 
($\Theta$). The horizontal line is the galactic plane; the rectangular boxes 
show schematically the Eastern (A) and Western (B) emission peaks of the CO 
ring (Fig.\,10). The shaded area shows the adopted idealized outflow cone
and its intersection with the galactic plane. The skew lines show 
'outstreamers, fingers, chimneys' of {\bf (a)} 8.4\,GHz continuum emission 
(Lisenfeld et al. 1996): heavy lines; dust fingers (Young et al.\ 1988, 
Hutchings et al. 1990): dashed lines; {\bf (b)} $^{12}$CO\,(1--0) (Young et 
al. 1988); {\bf (c)} $^{12}$\,CO\,(1--0) (this paper). The rotation of the 
galaxy is $\pm$\,{\bf 100}\,km\,s$^{-1}$ at $^{-}_{+}$\,1\,kpc radial 
distance. The numbers in the plot indicate the velocity (km\,s$^{-1}$) at 
which the outflows are detected (see also Fig.\,6).}
\end{figure}

a) {\it X--ray emission}. The ROSAT observations show X--\-ray emission 
extending out of the galactic plane as a feature called \mbox{{\it x--shaped}}
by Armus et al.\ (1995, their Fig.\,3). A tentative interpretation of the 
X--ray image is emission confined to a bi--conical outflow of opening angle 
$\Theta$ $\approx$ 60 -- 70$\degr$, pa $\approx$ 35$\degr$, and diameter at 
the base of {\it b} $\la$\,2\,kpc. Armus et al. mention that the X--ray 
emission is detected up to a height of $\sim$\,4\,kpc above the galactic plane.

b) {\it Synchrotron emission}. Interferometer observations between 327\,MHz and
15\,GHz by Lisenfeld et al. (1996) show essentially a bar--like distribution 
of the emission, but also extensions skew and perpendicular out of the galactic
plane, resembling galactic chimneys or galactic fountains. Fig.\,8\,a shows 
the position and position angles of these extensions seen at 8.4\,GHz in the 
observations by Lisenfeld et al. Similar chimneys were recently detected in 
M\,82 by Wills et al. (1999).

c) {\it Dust 'fingers'}. The optical images published by Young et al. (1988) 
and Hutchings et al. (1990) show several prominent dust 'fingers' extending 
skew to the SE away from the galactic plane. This dust may consist of cool 
material being dragged--out of the disk by the superwind of the starburst, 
rather than being only material of spiral arm III (de Vaucouleurs 1950, 
Benvenuti et al. 1975). The position and position angles of the dust 'fingers'
are shown in Fig.\,8\,a.

d) {\it Dragged--out molecular gas}. We believe, as explained above, that the 
velocity channel maps of the $^{12}$CO and $^{13}$CO lines (Young et al. 1988,
their Fig.\,2; Fig.\,6\,$\&$\,7) indicate mole\-cular gas being dragged out 
of the disk by the superwind of the starburst. From these observations we 
extracted the positions and position angles of the dragged--out molecular gas 
as shown in Fig.\,8\,b\,$\&$\,8\,c.

NGC\,2146 rotates anti--clockwise, i.e.\ from SE to NW. It is expected that 
the outflow material carries part of this rotation, hence angular momentum, 
out of the disk into the halo. To illustrate this effect (Fig.\,8\,c) we note 
that the dragged--out material which is oriented away from the galactic disk 
in the direction North and West is seen at similar positive velocities as the 
rotation of the NW part of the galaxy ({\it r} $\leq$ 0); the dragged--out 
material which is oriented in the direction East and South is seen at similar 
negative velocities as the rotation of the SE part of the galaxy (0 $\leq$ 
{\it r}). In the direction toward the center of the starburst the dragged--out
material is seen at small veloci\-ties and oriented nearly perpendicular
out of the galactic disk.

\section{The outflow}
We explain the features of Fig.\,1,\,2,\,3\,$\&$\,8 as a \mbox{bi--conical} 
outflow along the minor axis with opening angle $\Theta$\,$\approx$\,60$\degr$
and diameter at the base of {\it b} = 1.0 -- 1.5\,kpc, i.e. the same as the 
distance between the emission peaks of the CO ring (Table 2 below). The 
outflowing material is collimated by the material of the molecular ring. The 
geo\-metry of this outflow\footnote{for a similar figure of M\,82 see McKeith 
et al. (1995).} is schematically shown in Fig.\,9. In order to agree with the 
observed velocities of component I and II, in this interpretation the NE part 
of the galaxy is necessarily tilted toward us. Since the minor axis tilt 
$\epsilon$ = 90$\degr$ -- {\it i} and the half opening angle of the cone 
($\theta$ = $\Theta$/2) are {\it nearly} the same, i.e.\ $\epsilon$ $\approx$ 
$\theta$ $\approx$ 25 -- 30$\degr$, the rear side (I) of the NE cone and the 
front side (III) of the SW cone are seen {\it nearly} perpendicular to the 
line--of--sight so that the corresponding observed outflow velocity components
are small, or zero. These outflow components I and III are seen in the 
spectrogram Fig.\,1. The front side (II) of the NE cone and the rear side (IV)
of the SW cone are inclined by $\epsilon$ + $\theta$ $\approx$\,60\,$\degr$ to
the line--of--sight. For $\epsilon$ = $\theta$ = 30$\degr$ and a particular 
line--of--sight, the distances along the cone walls are
\begin{equation}
{\rm r}_{\rm a} = {\rm r(I)} = {\rm r(III)} = ({\it z} - 
{\it d})\,{\rm cos}({\epsilon}),\ 
\ \ ({\it b}/2)\,{\rm tan}({\epsilon})\ {\leq}\ {\it z}
\end{equation}
\begin{equation}
{\rm r}_{\rm b} = {\rm r(II)} = {\rm r(IV)} = [{\rm r}_{\rm a} + 
{\it b}\,{\rm sin}({\epsilon})]/{\rm cos}({\epsilon} + {\theta})
\end{equation}
with {\it z} the vertical height above the galactic plane and {\it d} the
semi--thickness of the galactic disk (0.3 -- 0.5\,kpc). Using the density 
relation n$_{\rm e}$({\it z}) of Eq.(1), the fact that the emission is 
proportional to n$_{\rm e}$$^{2}$, and the assumption that the cone walls have
a (constant) thickness L, the intensity $\cal I$ of the emission is
\begin{equation}
{\cal I}_{\rm a,b}\ \ {\propto}\ \ (n_{\rm e,a,b})^{2}\,{\rm L}\ 
{\propto}\ ({\rm r}_{\rm a,b})^{-4}\,{\rm L}
\end{equation}
The observed decrease of the H$_{\alpha}$ emission along the components I and 
III (i.e. $\cal I$$_{\rm a}$) shown in Fig.\,4\,a agrees reasonably well with 
this prediction. Since r$_{\rm b}$(II,IV) $\approx$ 3\,r$_{\rm a}$(I,III), we 
have $\cal I_{\rm b}$(II,IV) $\approx$ (1/100)\,$\cal I$$_{\rm a}$(I,III). 
This unfavourable intensity ratio explains why the components II and IV are 
not (easily) seen in a spectrogram taken along the minor axis (Fig.\,1). For 
this reason, in addition, we have calculated line ratios along the minor axis 
without taking into account possible line splitting (Fig.\,4\,$\&$\,5). 

\begin{figure*}
\vbox{
\psfig{figure=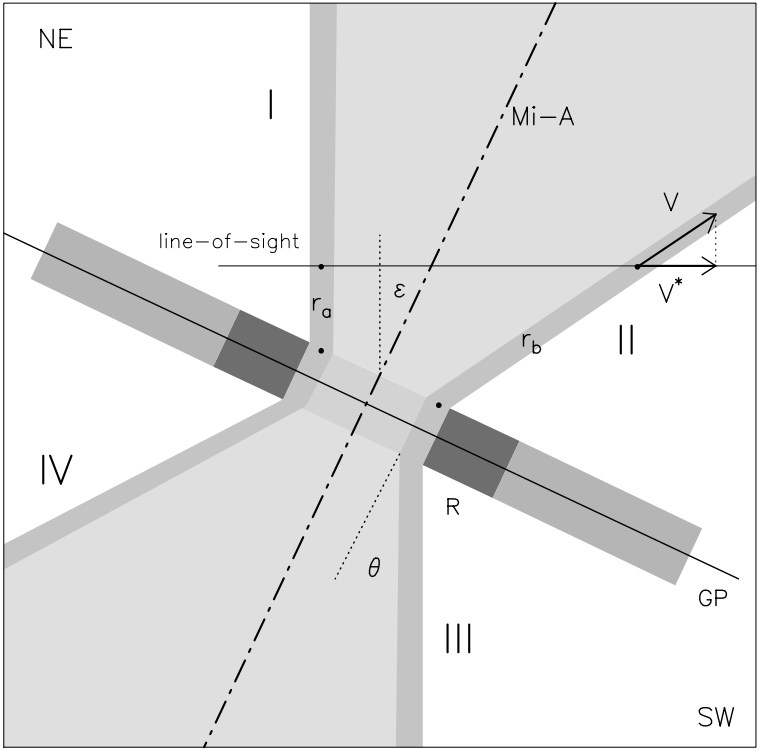,width=10cm,angle=0}\vspace{-5.0cm}}
\hfill\parbox[b]{7.5cm}{
\caption[]{Illustration of the outflow geometry of NGC\,2146. GP is the 
galactic plane, R the CO ring, Mi-A the minor axis, the observer is at the 
right side. The tilt of the minor axis is $\epsilon$ = 90$\degr$ -- {\it
i} with {\it i} = 60 -- 65$\degr$ the inclination of the galaxy. $\theta$ is
the half opening angle of the cones, i.e. $\epsilon$ $\approx$ $\theta$ 
$\approx$ 30$\degr$. The outflow cones, filled with X--ray emitting gas, and 
the cone walls \mbox{I -- IV}, emitting optical recombination lines, are shown. 
r$_{\rm a}$ and r$_{\rm b}$ are the distances along the cone walls from the 
galactic disk to the respective intersections with the line--of--sight. {\it V}
is the outflow velocity, {\it V}$^{*}$ the line--of--sight component. The 
diameter of the galactic disk, as shown here, is $\sim$\,6\,kpc.}
}
\end{figure*}

The observed velocity difference between \mbox{component I} and component II at
a height of 1 -- 2\,kpc NE above the galactic plane is $\Delta${\it v} 
$\approx$ 200\,km\,s$^{-1}$, with component II located at the blue side of 
component I (Fig.\,2\,$\&$\,3) in agreement with the geo\-metry of Fig.\,9. A 
similar value $\Delta${\it v} is reported by Armus et al. (1995), however also
for the SW side of the galactic plane. Since component I appears at a small 
velo\-city, we interpret $\Delta${\it v} $\approx$ {\it V}$^{*}$ as the 
blue--shifted line--of--sight velocity component of the outflow velocity 
{\it V} along cone wall II. From the geometry of the outflow we then obtain 
\mbox{{\it V} = {\it V}$^{*}$/sin($\epsilon$ + $\theta$)} $\approx$ 250 -- 
300\,km\,s$^{-1}$. This velocity is small compared to the terminal velocity of
$\sim$\,600\,km\,s$^{-1}$ predicted from the model by Chevalier $\&$ Clegg 
(1985), and compared to the outflow velocity of 500 -- 600\,km\,s$^{-1}$ 
measured in M\,82 (Table 2). This fact may contradict the tentative 
identification of component II. However, adopting a very different geometry to
avoid this situation would be in conflict with the data of Fig.\,8.  An 
explanation of this relatively low outflow velocity may be found in the much 
larger starburst volume of NGC\,2146, as compared to M\,82, the lower material
density in the starburst region, and in the apparently three times {\it smaller}
number of observed radio supernovae and supernova remnants in NGC\,2146 
(Tarchi et al. 2000), although the estimated star formation rate (Table 2
below), and hence supernova rate, seems to be three times higher. The larger 
volume, lower density, and the proportionally smaller number of radio 
supernovae and supernova remnants may produce in NGC\,2146 an outflow which is
less funneled because of the larger base and larger opening angle, and which 
is hence more diffuse and less propulsive. 

The conclusions reached above are based on an idealization of the outflow, 
which in reality may also be significantly distorted like the morpho\-logy of 
the galaxy. The present collection of observations, which contains some 
contradictions, does not allow a more precise determination of the actual 
situation.

\begin{figure*}
\vbox{
\psfig{figure=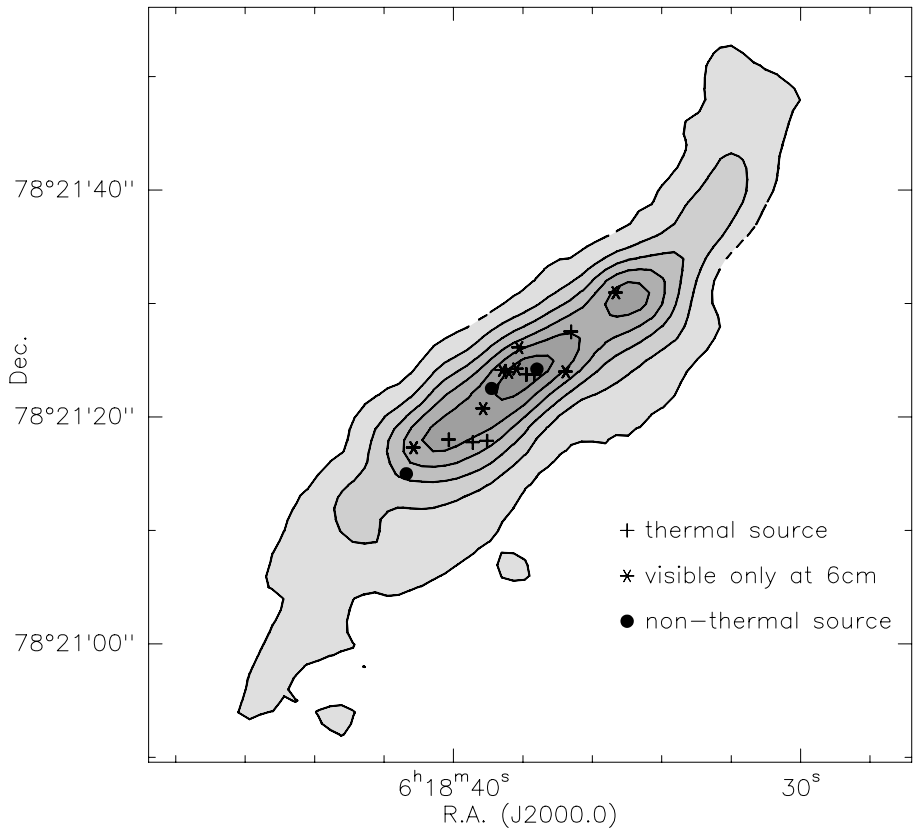,width=12.cm,angle=-90}\vspace{-2.0cm}}
\hfill\parbox[b]{5.5cm}{
\caption[]{Velocity--integrated $^{12}$CO \mbox{(1--0)} emission (Greve et al., 
in prep.) and radio sources observed with MERLIN and the VLA 
(Tarchi et al. 2000).}
}
\end{figure*} 

\begin{table*}
\caption[]{Minor Axis Outflows of NGC\,2146 and M\,82.}
\begin{center}\small
\begin{tabular}{lcc}
\hline
Parameter  & NGC\,2146$^{\ a)}$ & M\,82$^{\ a,b)}$ \\
\hline
Distance                   & 14.5\,Mpc & 3.2\,Mpc \\
{\bf The Starburst Region} &  & \\
Diameter starburst region & 2\,500\,pc & 850\,pc \\
Volume starburst region  & 2$\times$10$^{10}$\,pc$^{3}$ &
3$\times$10$^{8}$\,pc$^{3}$ \\
Diameter molecular ring (inside) & 750\,pc & 550\,pc \\
Radial thickness molecular ring ($\Delta${\it r}) & 1000/600\,pc$^{\ c)}$ 
& 250\,pc \\
Thickness molecular ring ($\Delta${\it z}) & 500\,pc & 150\,pc \\
Mass of starburst region & $\sim$\,1$\times$10$^{10}$\,M$_{\sun}$ & 
$\sim$\,2$\times$10$^{9}$\,M$_{\sun}$ \\
Density starburst region & $\sim$\,1/4\,M$_{\sun}$\,pc$^{-3}$ &
$\sim$\,3\,M$_{\sun}$\,pc$^{-3}$ \\
Infrared luminosity (L$_{\rm FIR}$)$^{\ d)}$ & 7.0
$\times$10$^{10}$\,L$_{\sun}$ & 2.4$\times$10$^{10}$\,L$_{\sun}$ \\
Star formation rate (SFR)$^{\ e)}$ & 5 -- 15\,M$_{\sun}$\,yr$^{-1}$ & 
2 -- 6\,M$_{\sun}$\,yr$^{-1}$ \\
  {\bf The Outflow}  &  & \\
Inclination galaxy ({\it i})$^{\ f)}$ & $\sim$\,60$\degr$ & $\sim$\,80$\degr$ \\
Visibility cones            & I,\,(II),\,III,\,{\bf --} & I,\,II,\,III,\,IV \\
Outflow base ({\it b}) & 1\,000\,pc & 500\,pc \\
Cone opening angle ($\Theta$\,=\,2\,$\theta$) & 60$\degr$ & 25$\degr$ \\
Outflow height ({\it h})  & $\sim$\,4\,kpc  & 2\,kpc (10\,kpc)$^{\ g)}$ \\
Volume of cones (up to {\it h}\,=\,3\,kpc) & 5$\times$10$^{10}$\,pc$^{3}$ 
& 1$\times$10$^{10}$\,pc$^{3}$ \\
Outflow velocity (at {\it h}\,$\approx$\,1\,kpc) & 250 -- 
300\,km\,s$^{-1}$ & 500 -- 600\,km\,s$^{-1}$ \\
Outflow acceleration & {\bf --} & 100 -- 150\,km\,s$^{-1}$\,kpc$^{-1}$ \\
Cone wall density (at {\it h}\,$\approx$\,1\,kpc) & $\sim$\,300\,cm$^{-3}$ & 
$\la$\,100\,cm$^{-3}$ \\ 
\hline
\multicolumn{3}{@{}l@{}}{%
a) Greve et al., in prep.}
\\
\multicolumn{3}{@{}l@{}}{%
b) Shopbell $\&$ Bland--Hawthorn (1998), southern cone; McKeith et al.
(1995).}
\\
\multicolumn{3}{@{}l@{}}{%
c) Eastern/Western section (see Fig.\,8\,$\&$\,10).}
\\
\multicolumn{3}{@{}l@{}}{%
d) L$_{\rm FIR}$ = 6$\times$10$^{5}$\,D$^{2}$\,[Mpc](2.58\,F$_{60}$[Jy] + 
F$_{100}$\,[Jy]) (Thronson $\&$ Telesco 1986).}
\\
\multicolumn{3}{@{}l@{}}{%
e) a best estimate from several authors using different methods.}
\\
\multicolumn{3}{@{}l@{}}{%
f) face--on: {\it i} = 0$\degr$.}
\\
\multicolumn{3}{@{}l@{}}{%
g) Devine $\&$ Bally (1999).}
\\
\end{tabular}
\end{center}
\end{table*} 

Figure 10 shows an overlay of the velocity--integrated $^{12}$CO\,(1--0) 
emission measured at PdB (Greve et al., in prep.) and the sources detected at 
1.6\,GHz and 5\,GHz (some) with MERLIN and the VLA (Tarchi et al. 2000). The 
sources are either radio supernovae, supernova remnants, 
ultra--compact/ultra--dense H\,{\sc ii} regions which contain massive star 
clusters or super star clusters. The objects are confined to the region in and
inside the CO ring, as expected from their origin in the starburst. This 
concentration to the CO ring and the inside provides additional evidence that 
supernova explosions and stellar winds are the primary source of the minor 
axis outflow. A similar figure for M\,82, based on PdB $^{13}$CO\,(1--0) 
observations and supernova observations, was published by Neininger et al.\ 
(1998, their Fig.\,3).

We believe that the ensemble of observations gives a consistent picture 
of an outflow and of dragged--out material, confined to a region of
$\sim$\,2\,kpc dia\-meter. Certainly, other explanations can be tested which
consider other known structures in and near the starburst region, in particular
a clumpy dusty spiral arm viewed under oblique angle and giving the impression
of dragged--out material. The dusty 'spiral arm III', proposed by de 
Vaucouleurs (1950), is probably a dust lane(s) with little mass since no 
trace is seen in deep dust continuum measurements (at 230\,GHz), and is located
at the edge of the starburst region as indicated by the visual obscuration. The
dust lanes are affected by the outflow, as we propose, but hardly affect the 
outflow. The incomplete string of H\,{\sc ii} regions, giving the impression 
of a spiral arm, is in fact an inclined $\sim$\,10$\times$20\,kpc ellipsoidal 
pattern partially projected onto the center region of the galaxy. This string 
of H\,{\sc ii} regions does not affect the outflow and is not affected by the 
outflow (Greve et al. in prep.). The 700\,pc long, triple--source 
{\bf S}--shaped pattern seen by Kronberg $\&$ Biermann (1981) at the center of
NGC\,2146 has been resolved into more sources hardly giving the impresssion of
a mini--spiral arm (Tarchi et al. 2000). On the other hand, the CO velocity 
pattern, in particular measured in the region 
$\sim$\,$\pm$\,800\,pc$\times$\,250\,pc along the major axis (i.e. inside the 
3rd contour in Fig.\,10), is that of a regularly rotating
disk and ring without perturbations. Above $\sim$\,300\,pc on either side of 
the galactic plane the velocity distribution flares out into the halo in a 
regular bi--cone like pattern (Greve et al., in prep.). The regularity of this 
velocity pattern in and above the galactic plane makes it difficult to 
envisage an alternative distribution of material and motions inside 
$\sim$\,2\,kpc diameter, which produces similar features as shown in Fig.\,8, 
than that of a starburst region with outflow. 
   
Table 2 compares the dimension of the starburst and the outflow in NGC\,2146 
with the prototype starburst and outflow in M\,82. The significantly larger 
geo\-metric scale and activity of the starburst phenomenon in NGC\,2146 is 
evident.

\section{Summary}
Using a combination of radio, optical, and X--ray observations, we have
tried to derive the geometry and physical state of an {\it idealized} minor 
axis outflow in NGC\,2146: the outflow is conical on either side of the 
galactic plane; the diameter at the base is {\it b} $\approx$ 1\,000\,pc, 
comparable in dia\-meter with the molecular ring; the opening angle is 
$\Theta$ $\approx$ 60$\degr$; near the base the hot outflowing material drags 
gas and dust out of the disk; the outflowing material carries angular momentum
into the halo; the cone walls of the outflow can be traced (at optical 
wavelengths) to a height of {\it h} $\approx$ 3\,kpc; the (electron) density 
in the cone walls seems to decrease proportional to {\it z}$^{-2}$, with 
{\it z} the height above the galactic plane; the outflow velocity at 
$\sim$\,1\,kpc height is 250 -- 300\,km\,s$^{-1}$.

While these parameters of the outflow give a consistent explanation of
the radio, optical, and X--ray observations, the absence of velocity
component IV and the low outflow velocity along the cone walls (at least
cone wall II) may speak against the proposed interpretation. However,
compared with M\,82, the larger dimension of the starburst region and 
of the molecular ring, and an apparently smaller number of radio 
supernovae and supernova remnants seem to produce a larger flaring of the
outflow ({\it b}, $\Theta$), and by this possibly a less propulsive 
outflow with a lower velocity. The weak component IV may also partially 
be hidden by the galactic disk, but its absence may also be in line 
with the general distortion of the galaxy. The coincidence of
ultra--compact/ultra--dense H\,{\sc ii} regions, radio supernovae, and
supernova remnants with the region of the CO ring and the interior
gives confidence in the picture of a superwind driven minor axis
outflow. \\

\begin{acknowledgements} We are thankful for the optical observations which 
were made in service time with the William Herschel Telescope (WHT) operated 
on the Island of La Palma by the INT Group in the Spanish Observatorio de 
Roque de los Muchachos of the Instituto de Astrofisica de Canarias. M. Bremer 
(IRAM) helped with the production of Figure\,1. We thank the referee for his 
comments and we agree with his remark that the presented picture is a tentative
interpretation of the sometimes scarce evidence of the outflow phenomenon. 
  
\end{acknowledgements}

\end{document}